\documentclass[prl,reprint,aps,amsmath,amssymb,twocolumn,superscriptaddress,longbibliography]{revtex4-1}

\usepackage{graphicx}% Include figure files
\usepackage{dcolumn}% Align table columns on decimal point
\usepackage{bm}% bold math
\usepackage{hyperref,xcolor,times}% add hypertext capabilities
\usepackage{siunitx}
\usepackage{ulem}
\usepackage{amsmath}
\usepackage{booktabs}
\usepackage{makecell}
\usepackage{enumitem}
\usepackage{float}
\usepackage{array}

\definecolor{linkblue}{rgb}{0, 0, 1}
\hypersetup{
	colorlinks=true,       		% false: boxed links; true: colored links
	linkcolor=linkblue,          	% color of internal links
	citecolor=linkblue,             % color of links to bibliography
	urlcolor=linkblue,          % color of external links
}

\newcommand{\ket}[1]{| #1 \rangle}
\newcommand{\bra}[1]{\rangle #1 |}

\begin{document}

\title{Ultrafast high-fidelity state readout of single neutral atom}

\author{Jian Wang}
\email{jwang28@ustc.edu.cn}
\affiliation{CAS Key Laboratory of Quantum Information, University of Science and Technology of China, Hefei 230026, China}
\affiliation{Anhui Province Key Laboratory of Quantum Network, University of Science and Technology of China, Hefei 230026, China}
\affiliation{CAS Center For Excellence in Quantum Information and Quantum Physics, University of Science and Technology of China, Hefei 230026, China}

\author{Dong-Yu Huang}
\thanks{These authors contributed equally to this work.}
\affiliation{CAS Key Laboratory of Quantum Information, University of Science and Technology of China, Hefei 230026, China}
\affiliation{Anhui Province Key Laboratory of Quantum Network, University of Science and Technology of China, Hefei 230026, China}
\affiliation{CAS Center For Excellence in Quantum Information and Quantum Physics, University of Science and Technology of China, Hefei 230026, China}
\affiliation{Hefei National Laboratory, University of Science and Technology of China, Hefei 230088, China}

\author{Xiao-Long Zhou} 
\affiliation{CAS Key Laboratory of Quantum Information, University of Science and Technology of China, Hefei 230026, China}
\affiliation{Anhui Province Key Laboratory of Quantum Network, University of Science and Technology of China, Hefei 230026, China}
\affiliation{CAS Center For Excellence in Quantum Information and Quantum Physics, University of Science and Technology of China, Hefei 230026, China}

\author{Ze-Min Shen}
\affiliation{CAS Key Laboratory of Quantum Information, University of Science and Technology of China, Hefei 230026, China}
\affiliation{Anhui Province Key Laboratory of Quantum Network, University of Science and Technology of China, Hefei 230026, China}
\affiliation{CAS Center For Excellence in Quantum Information and Quantum Physics, University of Science and Technology of China, Hefei 230026, China}

\author{Si-Jian He}
\affiliation{CAS Key Laboratory of Quantum Information, University of Science and Technology of China, Hefei 230026, China}
\affiliation{Anhui Province Key Laboratory of Quantum Network, University of Science and Technology of China, Hefei 230026, China}
\affiliation{CAS Center For Excellence in Quantum Information and Quantum Physics, University of Science and Technology of China, Hefei 230026, China}

\author{Qi-Yang Huang}
\affiliation{CAS Key Laboratory of Quantum Information, University of Science and Technology of China, Hefei 230026, China}
\affiliation{Anhui Province Key Laboratory of Quantum Network, University of Science and Technology of China, Hefei 230026, China}
\affiliation{CAS Center For Excellence in Quantum Information and Quantum Physics, University of Science and Technology of China, Hefei 230026, China}

\author{Yi-Jia Liu}
\affiliation{CAS Key Laboratory of Quantum Information, University of Science and Technology of China, Hefei 230026, China}
\affiliation{Anhui Province Key Laboratory of Quantum Network, University of Science and Technology of China, Hefei 230026, China}
\affiliation{CAS Center For Excellence in Quantum Information and Quantum Physics, University of Science and Technology of China, Hefei 230026, China}
\affiliation{Hefei National Laboratory, University of Science and Technology of China, Hefei 230088, China}

\author{Chuan-Feng Li}
\email{cfli@ustc.edu.cn}
\affiliation{CAS Key Laboratory of Quantum Information, University of Science and Technology of China, Hefei 230026, China}
\affiliation{Anhui Province Key Laboratory of Quantum Network, University of Science and Technology of China, Hefei 230026, China}
\affiliation{CAS Center For Excellence in Quantum Information and Quantum Physics, University of Science and Technology of China, Hefei 230026, China}
\affiliation{Hefei National Laboratory, University of Science and Technology of China, Hefei 230088, China}

\author{Guang-Can Guo}
\affiliation{CAS Key Laboratory of Quantum Information, University of Science and Technology of China, Hefei 230026, China}
\affiliation{Anhui Province Key Laboratory of Quantum Network, University of Science and Technology of China, Hefei 230026, China}
\affiliation{CAS Center For Excellence in Quantum Information and Quantum Physics, University of Science and Technology of China, Hefei 230026, China}
\affiliation{Hefei National Laboratory, University of Science and Technology of China, Hefei 230088, China}

\date{\today}
	
%\date{\vspace{0.1cm}\today}

\begin{abstract}

The capability to measure the state of a quantum system is vital to a practical quantum network, for applications including distributed quantum computing and long-distance quantum communication. As a thriving platform for quantum information technology, single neutral atoms suffer from low achievable photon scattering rate and shallow trapping potential, which limits the fidelity and speed of state readout process. Here, by coupling an single neutral atom with a high-finesse fiber-based Fabry-Pérot microcavity (FFPC) in Purcell regime, we realize strong enhancement of the atomic photoemission rate, which enables ultrafast and high-fidelity discrimination of bright and dark hyperfine states of the atom. The readout fidelity can reach 99.1(2)\% within 200 ns and 99.985(8)\% within 9 $\mu$s. Furthermore, we demonstrate that state preparation via optical pumping can be efficiently accelerated by real-time decision protocol based on ultrafast state readout. This work paves the way to the implementation of quantum networking protocols with high communication rate and high fidelity.

\end{abstract}

\maketitle
 
\renewcommand{\thefootnote}{}
\footnotetext[3]{These authors contributed equally to this work.}
%\titlespacing*{\section} {0pt}{3.5ex plus 1ex minus .2ex}{1ex plus .2ex}

% Introduction
The ability of full control over single quantum systems enables the construction of quantum networks and their vast applications \cite{kimble2008}. To tackle errors that stem from the fragility of qubits and quantum channels, quantum repeater protocols are proposed \cite{cz1998repeater}. They enable the scaling up of quantum networks by encoding logical qubits \cite{steane1996prl}, entanglement purification \cite{zeilinger2001purify} and entanglement swapping \cite{RMP2023repeater}, at the cost of additional qubit operations including state preparation, gate manipulation and state readout. Therefore, the speed and fidelity of quantum state preparation and readout along with basic gate manipulation, can determine the capability of entanglement generation between quantum network nodes \cite{rempe2015rmp, bernien2023npj}, as they are building blocks of elementary entanglement generation and logical qubit operations \cite{jiang2016scirep}. Furthermore, device-independent protocols \cite{weinfurter2022diqkd} and the test of quantum nonlocality \cite{hanson2015loophole, weinfurter2017loophole} rely on the space-like separation between operations of parties \cite{RMP2014bell}, which imposes stringent quantitative requirement on the state readout speed of quantum systems.

Owing to great scalability \cite{endres6100} and efficient photon interface \cite{rempe2007sci}, single neutral atom is a promising candidate for physical implementation of quantum network \cite{bernien2023npj}. Fast and high-fidelity single-qubit rotation and two-qubit gates have been demonstrated, which are fundamental building blocks of atom-based quantum networks \cite{lukin2024qec, kaist2018UltrafastRotation, NP2022ultrafastRydberg}. However, current technology of atomic state preparation and readout are still limiting the overall communication rate \cite{lucas2020entanglement} and fidelity \cite{rempe2021gate}. For example, state preparation and readout account for the majority of time consumption in atom-photon entanglement generation \cite{vv2024qecNetwork}. For free-space atoms, state-selective fluorescence detection \cite{lucas2008detection, bonn2017detection, weiss2019detection} and ionization-based state readout \cite{weinfurter2010ionization} have been achieved, resulting in the demonstration of mid-circuit measurement \cite{endres2023qec, thompson2023qec} and device-independent quantum key distribution \cite{weinfurter2017loophole, weinfurter2022diqkd}. Due to low photon collection efficiency, free-space fluorescence detection of tweezer-trapped atoms is difficult to reach a high fidelity with time consumption on the order of hundreds of microseconds. The strong coupling between atoms and optical cavity can vastly increase photon collection efficiency \cite{rempe2010cavDetect,rempe2014nat, S-K2022cavDetect, simon2024, vv5bit} to realize fast, high-fidelity and nondestructive state readout. And the atomic state can also be inferred by probing cavity reflection and transmission signal in strong coupling regime \cite{kimble2006PRL,orozco2009, LKB2010cavDetect}. However, practical quantum network applications still demand further advancements in nondestructive state readout methods with higher speed to enhance the communication rate \cite{vv2024qecNetwork, lanyon2023repeater, lucas2020entanglement} and higher fidelity to minimize the consumption of physical qubits \cite{knill2005, steane2007, RMP2023QEC}.

In this work, by coupling the atom and a fiber-based Fabry-Pérot microcavity (FFPC) in Purcell regime \cite{purcell1946} to simultaneously enhance the inherent atomic photoemission rate and collection efficiency, we realize ultrafast high-fidelity state readout of a single neutral atom. Strong Purcell enhancement leads to a photon detection rate of over 18 Mcps, which enables a readout fidelity of 99.1(2)\% within 200 ns and 99.985(8)\% within 9 $\mu$s. And the survival probability of the atom after each state readout process is over 99.7\%. With the aid of this state readout method which is much faster than typical optical pumping, we demonstrate its acceleration effect on atomic state preparation by applying a real-time decision protocol. This work demonstrates Purcell-enhanced state readout and preparation, and further highlights the application of the strong Purcell effect in advancing the operation of quantum networks.

\begin{figure}[tb]
    \centering \includegraphics[width=0.4\textwidth]{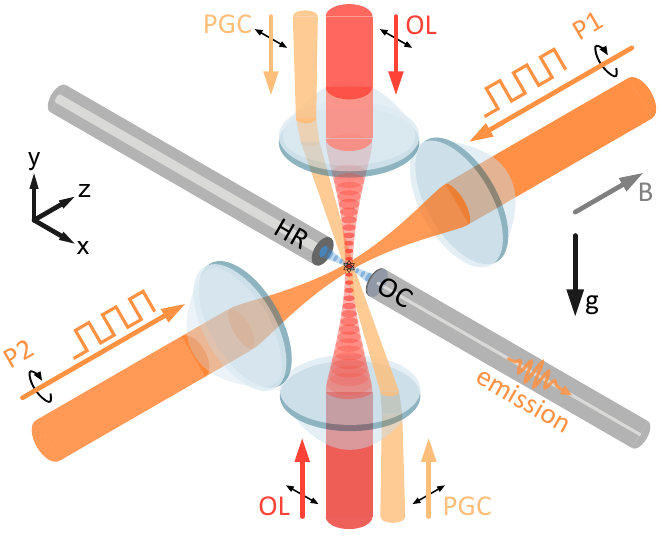}
\caption{Schematic diagram of the atom-FFPC system. At the center of the FFPC, the atom is trapped by a two-dimensional optical lattice. The two cavity mirrors are labeled as higher-reflective (HR) and out-coupling (OC). The square waveforms on probe beams indicate our applied alternate switching strategy to avoid interference while minimizing momentum transfer. For simplicity, laser beams for state preparation are not shown.
}
\label{setup} 
\end{figure}

Our experimental setup (Fig.\ref{setup}) contains a single $^{87}$Rb atom trapped by a two-dimensional optical lattice formed by two counter-propagating 850 nm laser beam (OL) and 776 nm intra-cavity standing wave. A pair of confocal aspherical lenses focus the 850 nm laser beams to a diameter of 11 $\mu$m. And a pair of linearly orthogonal-polarized counter-propagating 780 nm laser beams serve as polarization gradient cooling (PGC) laser. Along $z$-direction, we applied a bias magnetic field of 200 mGs as quantization axis. Another pair of aspherical lenses focus two counter-propagating 780 nm $\sigma_+$-polarized probe beams to a diameter of 20 $\mu$m. The two beams (P1 and P2) have balanced intensity and are spatially overlapped to minimize net momentum transfer to the atom when applying continuous probing in our state readout process. An alternate switching strategy \cite{sup} is applied to avoid detrimental interference.

Our FFPC consists of two highly reflective mirrors spaced by 80 $\mu$m, each formed by laser-ablated and dielectric coated optical fiber facet. The mirror with higher transmittance (labelled OC in Fig.\ref{setup}) serves as the out-coupling channel which is directly linked to our filter and detection setup \cite{sup}. Once the atom is cooled down and trapped by the 2D optical lattice, photons scattered by the atom can be collected by the cavity and detected with a single photon counting module (SPCM) whose output signal is analyzed by an FPGA-based controller to initiate the experimental sequences.

Fluorescence-based detection with high speed and fidelity relies on sufficient detected photons for the discrimination of bright and dark states. Therefore it is necessary to achieve high detected photon rate depending on overall photon collection and detection efficiency and photoemission rate of the atom. This can be realized by modifying the inherent photoemission properties of the atom by reaching Purcell regime of atom-cavity coupling \cite{purcell1946}, so that the atomic decay rate can be strongly enhanced compared to free-space situation, and the photons are efficiently collected and out-coupled without coherent energy exchange between the cavity and the atom \cite{rempe2008prl}. Our atom-FFPC system works entirely in Purcell regime with $\{g_0,\kappa,\gamma\} = 2\pi\times\{ 124, 159, 3.03 \}\text{ MHz}$ on $^{87}$Rb $D_2$ transition, where $g_0$ is the maximum atom-cavity coupling strength of $D_2$ line cycling transition obtained from geometry of the cavity, with $\kappa$ and $\gamma$ are the decay rates of the cavity field and atomic polarization respectively.

\begin{figure}[b]
    \centering 
        \includegraphics[width=0.45\textwidth]{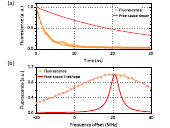}
    \caption{Characterization of atom-FFPC cooperativity. (a) Fluorescence lifetime measurement. An exponential fitting of the acquired fluorescence curve shows a $1/e$ lifetime of $2.51\pm 0.07$ ns, which is compared with the free-space decay curve (red) with a lifetime 26 ns. (b) Normalized fluorescence counts as a function of driving frequency with a Lorentzian fitting. The obtained full width half maximum (FWHM) of the curve is $62.46\pm 2.73$ MHz. A free-space curve is plotted for comparison. The origin of Frequency offset corresponds to the $^{87}\text{Rb}$ $\ket{^2S_{1/2},F=2}\rightarrow\ket{^2P_{3/2},F''=3}$ transition without a.c. Stark shift.
}
\label{measureC} 
\end{figure}

\begin{figure*}[t]
    \centering         \includegraphics[width=\textwidth]{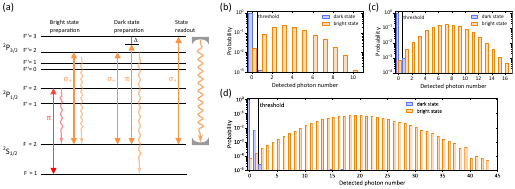}
    \caption{Laser configuration and results for characterization of state readout. (a) Lasers for state preparation and detection. For the preparation of bright state $\ket{F=2,m_F=2}$, a 795 nm laser resonant with $\ket{^2S_{1/2},F=1}\rightarrow\ket{^2P_{1/2},F'=2}$ transition pumps the atom out of $\ket{F=1}$. Simultaneously a 780 nm $\sigma_+$-polarized laser resonant with $\ket{^2S_{1/2},F=2}\rightarrow\ket{^2P_{3/2},F''=2}$ pumps the atom to $\ket{F=2,m_F=2}$. For dark state preparation, we apply the same $\sigma_+$ 780 nm laser to pump the atom to $\ket{F=1}$ and an additional 780 nm laser driving the $\ket{^2S_{1/2},F=2}\rightarrow\ket{^2P_{3/2},F''=3}$ transition with a detuning $\Delta = 80$ MHz to deplete atoms remaining in $\ket{F=2}$. Only atoms in bright state is resonantly driven and emit photons into FFPC in the state readout process. (b-d) Probability distribution of detected photon numbers in state readout processes with different duration $t_\text{R}$. (b) $t_\text{R} = 200$ ns. (c) $t_\text{R} = 800$ ns. (d) $t_\text{R} = 9\;\mu$s. All uncertainties are statistical. }
\label{hist} 
\end{figure*}

In the regime of strong Purcell enhancement with the cavity on resonance with the transition, the excited state lifetime is reduced by a factor of $(2C+1)$ where $C = g^2/(2\kappa \gamma)$ is the cooperativity of the system, and $g$ is the actual coupling strength between the atom and the cavity. This reduced lifetime can be extracted from the fast-decaying pattern of the fluorescence collected through the cavity \cite{lukin2014pcc}. Here, we send a 2 ns $\sigma_+$-polarized excitation laser pulse \cite{sup} that co-propagates with one of the probe beam. The pulse is on resonance with the transition $\ket{^2S_{1/2},\;F=2} \rightarrow \ket{^2P_{3/2},\;F''=3}$. The obtained fluorescence curve in Fig.\ref{measureC} (a) clearly shows a faster decay compared to free-space atoms. An exponential fit of the curve gives an enhancement factor of 10.45(29) and a cooperativity $C = $ 4.73(15). Measurement of the broadened transition linewidth also confirms the Purcell enhancement \cite{bonn2018purcell}. When applying continuous weak external driving with a Rabi frequency $\Omega$, the emission rate into the cavity
\begin{equation} 
R_\text{c} = \frac{\Omega^2}{\gamma C}\frac{|\Tilde{C}|^2}{|1+2\Tilde{C}|^2}, 
\end{equation}
where $\Tilde{C} = g^2/[2(\kappa-i\Delta_\text{c})(\gamma-i\Delta_\text{a})]$ is a modified complex cooperativity with $\Delta_{\text{a}/\text{c}}$ the detuning of the driving field with respect to the atom and cavity. Therefore the fluorescence lineshape is still Lorentzian but broadened by a factor of $(2C+1)$. We tune the driving frequency across the resonance to obtain the fluorescence counts corresponding to each laser frequency (Fig.\ref{measureC} (b)). The fitting shows a cooperativity $C = 4.65(23)$, which agrees with the lifetime measurement. The deviation of the measured $C$ from theoretical value of $g_0^2/(2\kappa \gamma) = 16$ mainly results from the polarization mismatch between the driven transition and the cavity mode which gives a factor of 1/2 \cite{Lanyon2023PRXQ}, and a spatial-averaged coupling strength experienced by the atom with finite temperature.

Next, the characterization of the fluorescence-based state readout is carried out. The experimental sequence is repeated as follows. The atom is first cooled for 1 ms, then it is prepared by optical pumping for 2 $\mu$s to bright state $\ket{F=2,m_F=2}$ or 25 $\mu$s to dark state $\ket{F=1}$ as described in Fig.\ref{hist} (a) alternately, followed by an additional 4-$\mu$s state readout pulse to confirm the success of state preparation. Then a state readout process of certain configuration is applied and the detected photon number $N$ is analyzed as in Fig.\ref{hist} (b-e). For each state readout configuration, an optimized discrimination threshold $N_\text{thr}$ is chosen, so that an inferred infidelity
\begin{equation}
\epsilon = \frac{ P(N\geq N_\text{thr}|\text{dark state}) + P(N< N_\text{thr}|\text{bright state})}{2}
\end{equation}
is minimized.

The choice of the bright state $\ket{F=2,m_F=2}$ along with the $\sigma_+$-polarized probe light makes the cycling transition $\ket{F=2,m_F=2}\rightarrow\ket{F''=3,m_F''=3}$ become the only one that is driven. Therefore, off-resonant scattering through the $\ket{F'=2}$ states are suppressed to avoid pumping the bright state atoms to the dark state, which allows the atom to emit sufficient photons for state readout. Meanwhile, the counter-propagating probe beam with switching strategy \cite{sup} minimizes the heating effect of the atom during state readout. The probability distribution of detected photon number for a probe duration $T=200$ ns and intensity of $587\text{ mW}/\text{cm}^2$ is shown in Fig.\ref{hist} (b). The distinction between distributions correspond to bright and dark states is clear despite the short readout duration, giving a fidelity of 99.1(2)\%. This is achieved by a high photon detection rate of 18.1 Mcps which is inferred from the distribution, with the error of dark state is composed by fiber Raman noise of 0.7 kps and stray photons from probe laser of 5.8 kps. With longer probe time and modest probe intensity, a fidelity of 99.985(8)\% and 99.91(3)\% can be achieved within 9 $\mu$s and 800 ns respectively as shown in Fig.\ref{hist} (c,d). For atoms in bright state, the survival probability after scattering the probe laser is measured to be 99.73(4)\%, 99.82(3)\% and 99.87(2)\% for readout time of 200 ns, 800 ns and 9 $\mu$s respectively.

It should be emphasized that the count rates inferred from distribution diagrams in Fig.\ref{hist} are at Mcps scale, which approaches the maximum value set by the SPCM dead-time of 28 ns, especially for the configuration with 200 ns readout time. This causes severe underestimation of photon numbers \cite{detectorNP}, which in fact "misses" around 50\% of photons in the 200 ns configuration. Taken this effect into consideration, an estimated overall detection efficiency of a photon in cavity mode is 26\% \cite{sup}. The underestimation of photon number due to dead-time, however, does not considerably compromise the state readout fidelity since one or two photons detected is sufficient for the present discrimination threshold.

\begin{figure}[t]
    \centering
    \includegraphics[width=0.45\textwidth]{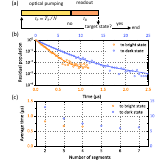}    
    \caption{Accelerated state preparation enabled by ultrafast state readout. (a) Experimental protocol. Each state readout is followed by a real-time decision on whether repeating the optical pumping or finishing the protocol. The repetition number of each run is recorded to obtain $\langle t \rangle$. (b) Residual population as a function for each optical pumping scheme. The $T_\text{P}$s for dark and bright states is set to be 24 $\mu$s and 1.0 $\mu$s respectively. (c) Average time for completing the protocol corresponding to different numbers of segments.}
\label{prep} 
\end{figure}

As recent work have demonstrated the state preparation can benefit from fast and high-fidelity state readout \cite{Wrachtrup2024SiC, Pan2024SiC}, we further demonstrate an acceleration effect on state preparation by performing the present state readout methods with real-time decision. Typical state preparation for atoms rely on optical pumping process, and the residual population follows an exponential decay pattern in time domain as shown in Fig.\ref{prep} (b). In this way, it takes a long optical pumping duration $T_\text{P}$ to ensure the success of state preparation, e.g. seven times of the $1/e$ lifetime to ensure a residual population less than $10^{-3}$. The accelerated-preparation protocol is straightforward, as shown in Fig.\ref{prep}. (a). The nature of exponential decay indicates that most of the atoms are transferred to the target state much earlier than reaching $T_\text{P}$. Therefore, if we break the long optical pumping process into $N$ segments, and perform a state readout after each segment to collapse the atom into the target state at the $i$-th segment with probability $P_\text{end}(i)$, then the average total duration of optical pumping can be greatly shortened. Suppose that $t_\text{R}$ is the duration of state readout, the average length of the protocol is
\begin{equation} 
\langle t \rangle = \sum_{i=1}^{n} P_\text{end}(i)i(t_\text{P}+t_\text{R}), \end{equation}
where $t_\text{P}=T_\text{P}/N$ is the length of an optical pumping segment. A detailed numerical simulation of $\langle t\rangle$ is presented in \cite{sup}. This protocol is, however, not favorable for former experiments on single neutral atoms, since a practical $T_\text{P}$ is at the order of several or tens of microseconds, which is comparable or even shorter than typical state readout processes. Here, given by $t_\text{R}$ much shorter than $T_\text{P}$, acceleration of state preparation is possible by applying this protocol.

For the preparation of dark state, we divide $T_\text{P} = 24\;\mu$s into several segments each followed by an $800$-ns state readout. An average time $\langle t\rangle$ for completing the protocol is minimized to 5.98 $\mu$s when $t_\text{P} = T_\text{P}/6 = 4\;\mu$s. For the bright state preparation where the pumping dynamics is much faster, readout pulses with a duration of 300 ns is applied. In this case, we obtain a $\langle t\rangle$ of 0.65 $\mu$s with the original $T_\text{P} = 1.0\;\mu$s when segment number $N=4$, corresponding to speed-up factors of 4.0 and 1.5 for preparation of dark and bright states respectively. Here, we deliberately apply a cutoff on the number of the "pumping-readout" cycles when it reaches $N$ for simplicity, and mark the trials in which the final state readout indicates a non-target state as failures. The proportion of failures is measured to be below 0.05\% and 0.3\% for preparation of dark and bright states, which is below the infidelity of the corresponding state readout.

In this work, based on atom-FFPC setup, we utilize the strong Purcell effect to simultaneously enhance the inherent photoemission rate and photon collection efficiency of a single neutral atom to perform ultrafast and high-fidelity state readout. The photon detection rate of more than 18 Mcps allows the discrimination of bright and dark states in 200 ns with infidelity below one percent. Higher fidelity of 99.91(3)\% and 99.985(8)\% is achieved with a readout time of 800 ns and 9 $\mu$s respectively. The atom loss probability after each readout of bright state is maintained to be below 0.3\%. With the readout speed surpassing those of traditional state preparation, we demonstrate an accelerated state preparation with the aid of ultrafast state readout, which implies the cooperation between different stages of qubit operation and can promote the understanding of quantum protocol design to optimize the performance of a quantum platform. As the overall speed and fidelity of quantum networking protocols can be significantly benefit from the present state readout method, this work paves the way to the realization of scalable and high-performance quantum network \cite{vv2024qecNetwork}.

The speed and fidelity of the presented readout method in Purcell regime can be further improved. By performing three-dimensional tight confinement and ground state cooling, the cooperativity of the atom-FFPC system can be stabilized at a larger value. Better FFPC-fiber coupling and more powerful photon detector can also lead to a higher detection efficiency. These lead to a shortened readout time and higher fidelity.

Benefit from the large optical access of FFPC, the present work can be extended to atom arrays \cite{satellite} to perform local two-qubit entangling gate and mid-circuit measurement. Recent work \cite{lukinffpc} have demonstrated the ability of integrating two tweezer-trapped atoms to a single FFPC and performing fast state readout and entangling gate. Applying dark state encoding \cite{weinfurter2017loophole} or site-selective level shifting \cite{vv5bit, satellite}, Purcell-enhanced fluorescence readout can be performed on a single atom without affecting the rest.

\begin{acknowledgments}
This work was supported by the Innovation Program for Quantum Science and Technology (No.\,2021ZD0301200), the National Natural Science Foundation of China (No.\,11804330 and No.\,11821404), and the Fundamental Research Funds for the Central Universities (WK2470000038).
\end{acknowledgments}

\vspace{0.6cm}

\bibliography{ref.bib}

\newpage
\appendix
\setcounterref{figure}{0}

\section{\Large{Supplementary Meterials}}
\subsection{Single atom preparation}

\subsubsection{Loading of single atom}
To load the single neutral atom in the two-dimensional optical lattice, we apply an experimental sequence as described below. The procedure starts with the preparation of a magneto-optical trap (MOT) 2 mm above the fiber-based Fabry-Pérot microcavity (FFPC) for 3 s followed by a 2-ms long polarization gradient cooling. With sufficient number of atoms being cooled and 1 s before the MOT is released, a 796 nm guiding beam is turned on to reduce the spread of the atoms in the horizontal plane during the falling. Then the atoms are decelerated by the PGC beams described in main text and finally trapped by the two-dimensional optical lattice. The mean number of the trapped atoms are maintained sufficiently low so that only one atom is trapped in each run of experiments.
\begin{figure}[h]
	\centering
	\includegraphics[width=0.4\textwidth]{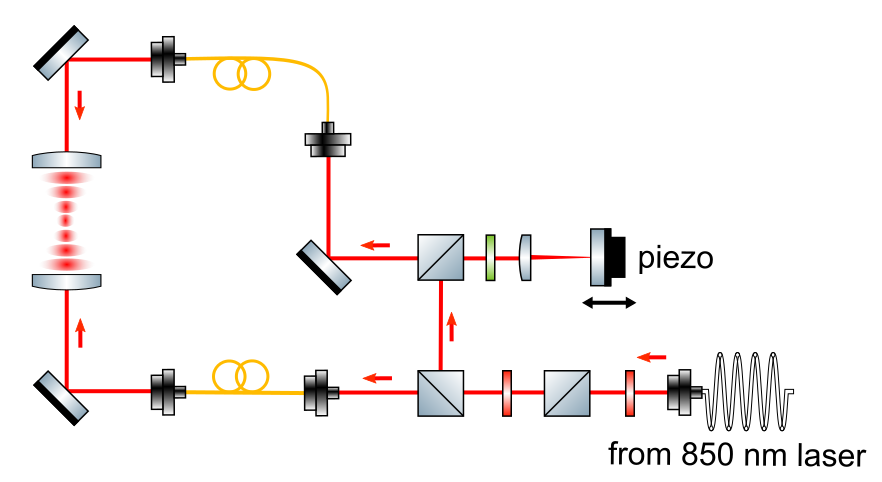}
	\caption{Setup for atom position optimization by adjusting the phase of the 850 nm optical lattice. A mirror attached to a stack piezoelectric actuator is applied to change the relative phase of the two laser beams. Optical components can be referred to in Fig. \ref{776}.}
	\label{850}
\end{figure}
\subsubsection{Optimization of position}
Once the Rb atom is trapped inside the FFPC, which is indicated by a significant increase in the photon counting signal from the SPCM (SPCM-AQRH-54-FC, Excelitas), the 796 nm guiding beam is turned off and the optimization process of atom position is initiated. This is done by adjusting the relative phase of two 850 nm laser beams that form the red-detuned optical lattice. The details of the setup is shown in Fig. \ref{850}. The phase is controlled by a piezoelectric module whose voltage is scanned with a triangle wave. Once the counting rate of the scattered photons reaches a preset threshold, the scanning stops and the atom's position is estimated to be optimized.

\begin{figure}[h]
	\centering
	\includegraphics[width=0.5\textwidth]{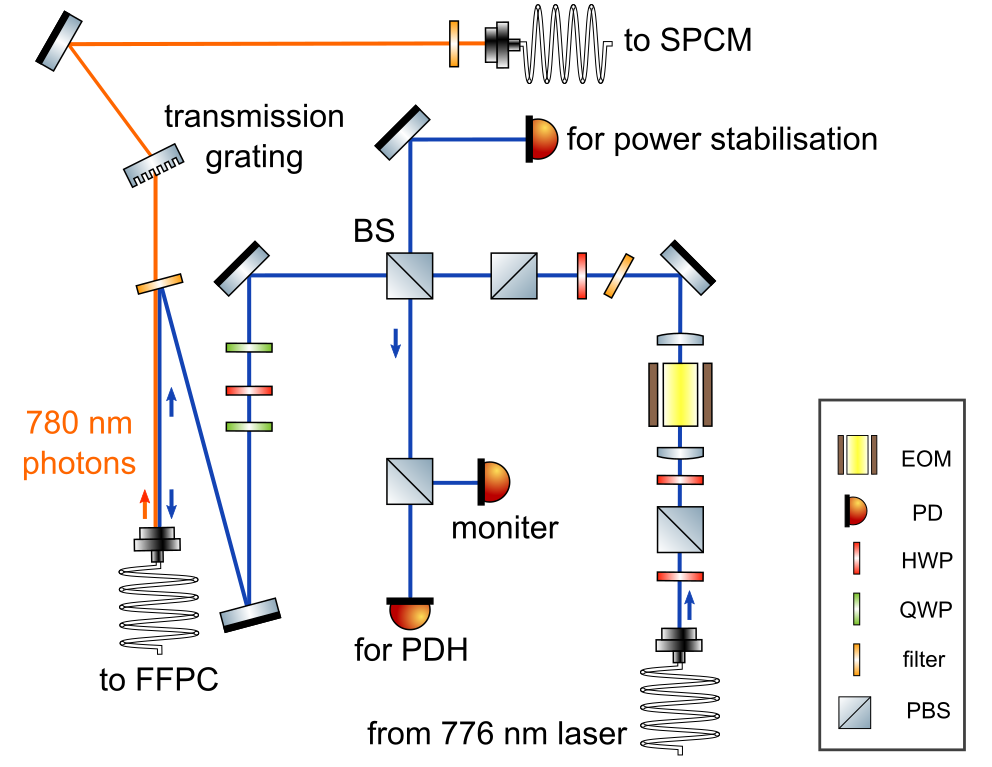}
	\caption{Setup for cavity stabilization and fluorescence detection. A free space EOM modulates the 776 nm laser at a frequency of 700 MHz to generate the PDH error signal for FFPC stabilization. The out-coupled 780 nm photons are separated from the 776 nm beam path with a narrow line optical filter. Abbreviations of optical components are listed below, EOM: electro-optic modulator, PD: photodetector, HWP: half waveplate, QWP: quarter waveplate, filter: 780 nm narrow-line filter, PBS: polarizing beam splitter.}
	\label{776}
\end{figure}

\subsection{Cavity stabilization and fluorescence detection setup}

To maintain resonant with the atomic transition, our FFPC is actively stabilized by a 776 nm laser with Pound-Drever-Hall (PDH) technology \cite{pan2022oe} which also serves as intra-cavity optical lattice. The frequency of the laser is one free spectral range (FSR) away from the $^{87}\text{Rb}$ $\ket{^2S_{1/2},F=2}\rightarrow\ket{^2P_{3/2},F''=3}$ transition, and it is actively stabilized by beating with an optical frequency comb (FC1500-250-ULN, Menlo Systems) \cite{shen2023rsi}. Since the 776 nm laser is close to the wavelength of atomic fluorescence, the PDH setup causes large amount of noise in our photon detection, mainly due to the Raman scattering \cite{bonn2016apb} of the intense laser in optical fibers. Therefore special cares should be taken in the photon detection setup. As shown in Fig. \ref{776}, we use a transmission grating (T-1400-800-12.7*12.7-94, LightSmyth) and several narrow-line optical filters (LL01-780-12.5, Semrock) to reduce the noise. A final noise of 0.7 kcps is obtained which is sufficient for our state readout experiments.

\subsection{Probe laser setup}

\subsubsection{Fast excitation with short laser pulse}
\begin{figure}[h]
	\centering
	\includegraphics[width=0.5\textwidth]{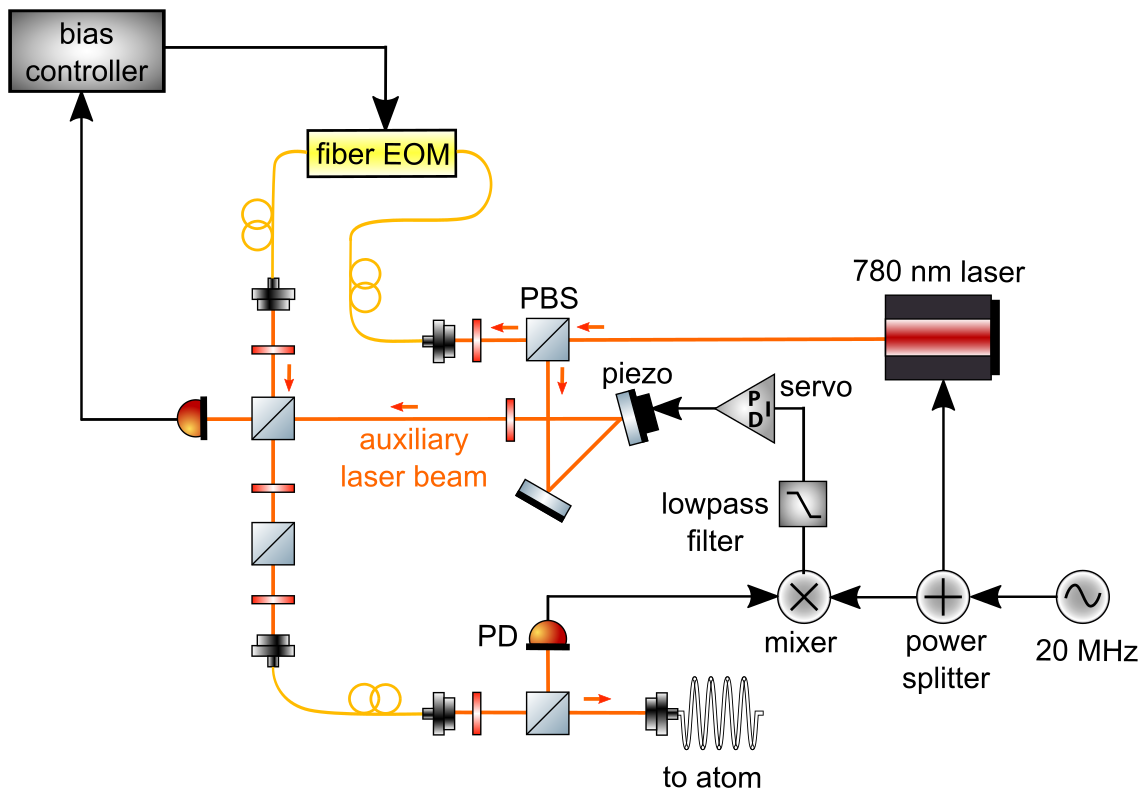}
	\caption{Setup for active background cancellation for fast excitation of the atom. A bias controller is applied to stabilize voltage of the fiber EOM prior to the interference. The interferometer is stabilized by demodulating the monitor signal with a modulation of the laser. Optical components can be referred to in Fig. \ref{776}.}
	\label{2ns780}
\end{figure}

\begin{figure}[h]
	\centering
	\includegraphics[width=0.4\textwidth]{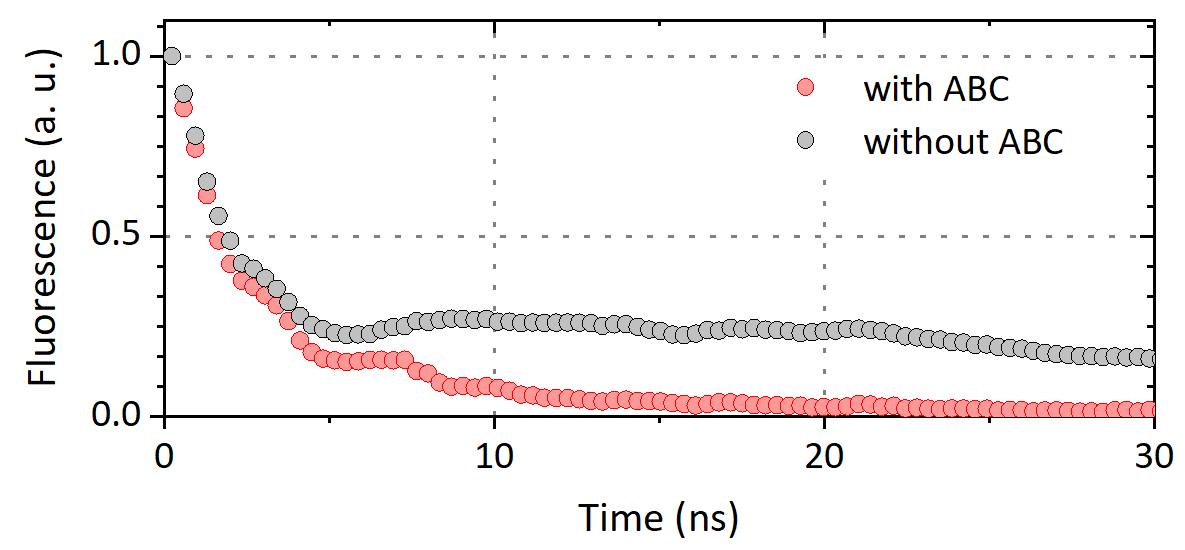}
	\caption{Comparison of the fluorescence curve with and without active background cancellation.}
	\label{blur}
\end{figure}

The measurement of the fluorescence curve of the atom-FFPC system requires a fast laser pulse for atomic excitation. The pulse should be turned off fast enough compared to the decay of the atom. Therefore we apply an intensity fiber electro-optic modulator (fiber EOM, NIR-MX800-LN-10, iXblue) to generate the 2-ns short laser pulse. However, commercial modulators suffer from a finite on-off ratio which results in continuous excitation that severely blurs the fluorescence curve as in Fig. \ref{blur}. To mitigate this issue, we employ an active background cancellation technique to improve the on-off ratio as describe in Fig. \ref{2ns780}. This is achieved by the interference between the modulated laser and an auxiliary laser beam whose intensity is equal to the background intensity of the modulated one. The phase of the interferometer is set to make the background laser field destructively interferes with the auxiliary laser and the laser pulse itself is almost unaffected. The obtained fluorescence curve using background cancellation technique is clearly improved.


\subsubsection{Continuous laser probing}
Ultrafast and high-fidelity state readout of the atom requires the scattering of enough photons in a short readout duration. Therefore, to minimize the heating effect due to photon absorption and recoil, it is favorable to use two counter-propagating laser beams to drive the atomic transition. A typical configuration is to use a pair of lin$\perp$lin laser beams to avoid interference. This is, however, not suitable in our experiments since we require a purely $\sigma_+$ light field to drive the only transition $\ket{F=2,m_F=2}\rightarrow\ket{F''=3,m_F''=3}$ and avoid scattering through the $\ket{F''=2}$ states. Therefore, we apply an alternate switching strategy as follows. We drive the acousto-optic modulator (AOM) that switch on the probe laser with a square waveform to get a pulse train with a pulse width of 50 ns and period of 100 ns. Then, the laser beam splits into two paths with one directly enter the vacuum chamber (as P1 in main text), while the other is delayed by a 10-m optical fiber and enter the chamber from the opposite direction (P2). The two paths share the same intensity and same rotating direction of electric field vectors. In this way, the atom is driven by the two beams in an alternate mode with only the $\sigma_+$ cycling transition driven and the net momentum transfer to the atom is cancelled.

\subsection{Numerical simulation of accelerated state preparation}

\begin{figure}
	\centering
	\includegraphics[width=0.4\textwidth]{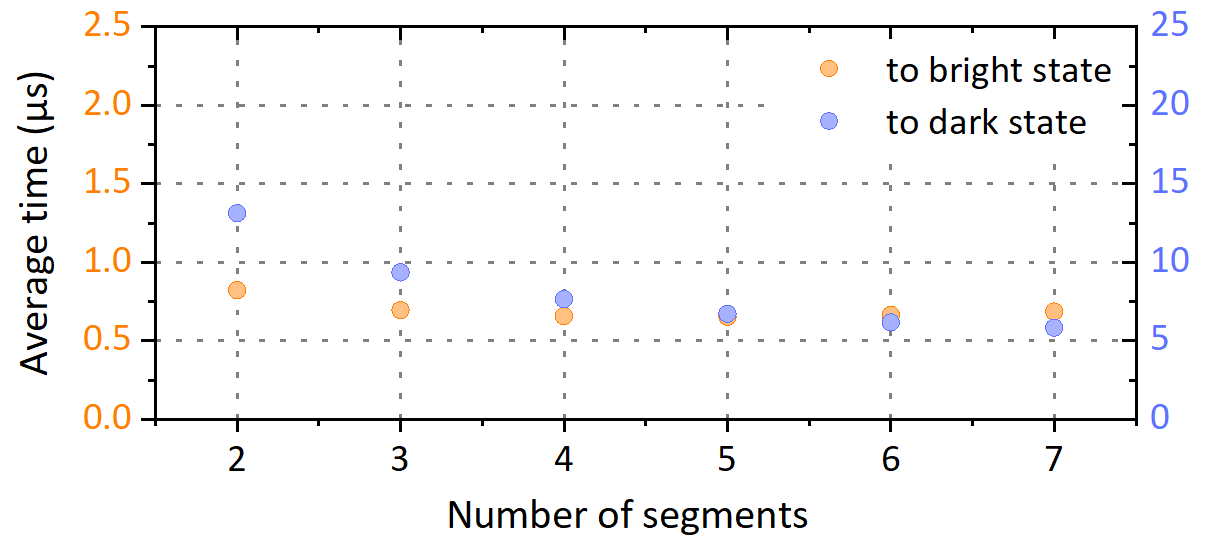}
	\caption{A numerically simulated $\langle t\rangle$ of different configurations in the accelerated state preparation. }
	\label{ASP}
\end{figure}

The accelerated state preparation protocol can be analyzed in the formalism of transition matrix. As the processes are incoherent, the state of the atom can be described by a two-dimensional real vector $s = (p_\text{i}\;p_\text{t})^\text{T} $ as the components are the probabilities of the atom in initial and target states respectively. Suppose that the residual population of the optical pumping process follows an exponential decay pattern with a time constant $\tau$ and steady state value $r$, the transfer matrix of the optical pumping can be expressed as 
\begin{equation}
	M_\text{P}(t) = \begin{pmatrix}
		r+(1-r)e^{-t/\tau} & re^{-t/\tau}\\
		(1-r)(1-e^{-t/\tau}) & 1-re^{-t/\tau}
	\end{pmatrix}.
\end{equation}
Suppose that $s_i^0$ is the initial state before the $i$-th run of the protocol, then after the optical pumping with a duration $t_\text{P}$, the state becomes
\begin{equation}
	s_i^1 = M_\text{P}.
\end{equation}
Therefore the probability of finishing the protocol right here $P_\text{end}(i)$ is the second component of $s_i^1$ multiplied by the product of all the first components of $s_j^1$ with $j<i$, since this event can be interpreted as "the atom is in initial state from trial $1$ to $(i-1)$" AND "the atom is in target state in trial $i$". In this way, the average time $\langle t \rangle$ can be obtained with the formula in the main text. A numerically simulated $\langle t \rangle$ with experimental parameters from the main text is presented in Fig. \ref{ASP}.

\subsection{Analysis on photon emission rate}

The interaction between a laser-driven atom and an optical cavity can be described by Lindblad equation with the following coherent Hamiltonian and dissipative jump operators \cite{rempe2015rmp}:
\begin{align}
	H =& -\Delta_{a}\ket{e}\bra{e} - \Delta_c a^\dagger a  \notag\\ &+ g(a\ket{e}\bra{g} + a^\dagger\ket{g}\bra{e}) + \frac{\Omega}{2}(\ket{e}\bra{g} + \ket{g}\bra{e}), \\
	L_1 =& \sqrt{2\gamma}\ket{g}\bra{e}, \\
	L_2 =& \sqrt{2\kappa}a,
\end{align}
where $\ket{g}$ and $\ket{e}$ denote the ground and excited states of the atom, and $a$ is the annihilation operator of the cavity field. The detuning of the atom (cavity) with respect to the driving laser is labelled $\Delta_{a(c)}$ and $\Omega$ is the Rabi frequency of the laser. In our state readout experiments, intense driving is applied for strong emission, so that low saturation approximation is no longer appropriate. Therefore, we apply Qutip \cite{qutip} to numerically solve the Lindblad equation for a theoretical photon emission rate.

After solving for the steady state solution, a photon emission rate into the cavity $R_\text{c} = 2\kappa\langle a^\dagger a\rangle $ can be obtained, and the result corresponding to each state readout configuration is listed in Table. \ref{rate}. The detection rates $R_\text{d}$ inferred from the photon number probability distribution is also listed, along with the dead-time-corrected value which is obtained by multiplying a factor of $(1-R_\text{output}t_\text{dead})^{-1}$. Therefore an estimated overall efficiency of detecting an emitted photon is 26\%, which is the product of the listed factors:
\begin{itemize}[itemsep=0.5pt, topsep=0.5pt, parsep=0.5pt,leftmargin=15pt]
	\item probability of the cavity photon leaking through the out-coupling mirror,
	\item coupling efficiency from the cavity mode to the fiber guide mode,
	\item overall transmittance of all the optics in the filter setup $\sim 90\%$,
	\item coupling efficiency of the fiber connected to the SPCM $\sim 85 \%$,
	\item quantum efficiency (QE) of the SPCM $\sim 65\%$.
\end{itemize}

The first two factors are compromised compared to our designed value, due to the slight misalignment of the FFPC during the vacuum bake-out process. Therefore a significant improvement in the overall detection efficiency is feasible by improving the thermal stability of the FFPC, optimizing the filter setup and applying superconducting nanowire single photon detector whose QE can reach over 90\%.\\

\begin{table}[H]
	\centering
	\setlength{\tabcolsep}{8pt}
	\begin{tabular}{cccc}
		\hline
		\thead{Readout\\ time (ns)} & \thead{Calculated \\ $R_\text{c}$ (Mcps)} & \thead{Photon detection\\ rate $R_\text{d}$ (Mcps)} & \thead{Dead-time-corrected\\ detection rate (Mcps)}\\
		\hline
		200 & 140 & 18.1 & 36.8\\
		800 & 47 & 9.12 & 12.3\\
		9000 & 8.8 & 2.16 & 2.29\\
		\hline
	\end{tabular}
	\caption{Photon emission rate in different experimental configurations.}
	\label{rate}
\end{table}

\end{document}